\documentclass{ecai}
\usepackage{times}
\usepackage{graphicx}
\usepackage{latexsym}
\usepackage{xcolor}

\makeatletter  
\newif\if@restonecol  
\makeatother

\usepackage[linesnumbered,ruled,vlined]{algorithm2e}
\usepackage{algpseudocode}  
\usepackage{multirow}

\usepackage{subfig}
\usepackage{enumerate}

\usepackage{amsthm,amsmath,amssymb}
\newtheorem{Def}{Definition}
\usepackage{mathrsfs}

\begin{document}

\title{Private Knowledge Transfer via Model Distillation with Generative Adversarial Networks}

\author{Di Gao \and  Cheng Zhuo\institute{Zhejiang University,
China, email: czhuo@zju.edu.cn} }

\maketitle

\begin{abstract}
The deployment of deep learning applications has to address the growing privacy concerns when using private and sensitive data for training. A conventional deep learning model is prone to privacy attacks that can recover the sensitive information of individuals from either model parameters or accesses to the target model. Recently, differential privacy that offers provable privacy guarantees has been proposed to train neural networks in a privacy-preserving manner to protect training data. However, many approaches tend to provide the worst case privacy guarantees for model publishing, inevitably impairing the accuracy of the trained models. 
In this paper, we present a novel private knowledge transfer strategy, where the private teacher trained on sensitive data is not publicly accessible but teaches a student to be publicly released. In particular, a three-player (teacher-student-discriminator) learning framework is proposed to achieve trade-off between utility and privacy, where the student acquires the distilled knowledge from the teacher and is trained with the discriminator to generate similar outputs as the teacher. We then integrate a differential privacy protection mechanism into the learning procedure, which enables a rigorous privacy budget for the training. 
%The proposed framework employs adversarial learning to distill knowledge and transfer between teacher and student, and then integrates a privacy-preserving learning procedure to allow student training using only unlabelled public data and very few epochs. 
The framework eventually allows student to be trained with only unlabelled public data and very few epochs, and hence prevents the exposure of sensitive training data, while ensuring model utility with a modest privacy budget. The experiments on MNIST, SVHN and CIFAR-10 datasets show that our students obtain the accuracy losses $w.r.t$ teachers of 0.89\%, 2.29\%, 5.16\%, respectively with the privacy bounds of (1.93, $10^{-5}$), (5.02, $10^{-6}$), (8.81, $10^{-6}$). When compared with the existing works \cite{papernot2016semi,wang2019private}, the proposed work can achieve 5-82\% accuracy loss improvement. 

%PATE \cite{} reports 1.21\%, 10.88\% accuracy losses on MNIST and SVHN and RONA \cite{wang2019private} reports 8.98\% accuracy loss on CIFAR under the same privacy bound.
\end{abstract}

\section{INTRODUCTION}

At the era of big data, the recent breakthroughs of computing infrastructures and neural network algorithms have facilitated the adoption of deep learning in various domains from facial recognition to health care management. The successful deep learning applications in real-world services depend on not only the high-performance inference models but also the quantity and the quality of training data. 

%\textbf{Privacy concerns in deep learning.}
Such training data often contains private and sensitive information, $e.g.$, facial features, financial records, health history, $etc.$, inevitably causing the risk of privacy leakage for the data owners. Thus, there have been increasing privacy concerns with the growing deployment of deep learning applications. Since deep neural network itself can function as an encoder by translating the individual data into model parameters, many prior works \cite{fredrikson2015model,shokri2017membership} have demonstrated the 'hacking' of sensitive information from the neural network, the performance of which can be notably improved if the attacker can repeatedly query outputs of the model even in a 'black-box' manner \cite{tramer2016stealing}. Thus, it is highly desired to {have privacy-preserving techniques for deep learning applications that can ensure model utility while protect sensitive information}. 
%a strong attacker can achieve model-inversion in 'black-box' access manner in which model parameters are not invisible. 

%\textbf{Techniques of privacy protection}
Recent researches \cite{abadi2016deep,shokri2015privacy,yu2019differentially} have investigated privacy protection from various aspects. The concept of privacy-preserving deep learning was first proposed in \cite{shokri2015privacy}, in which multiple private participants jointly trained a model by updating the sanitized model parameters, while the training data were kept at local. A more general approach was then proposed by \cite{abadi2016deep} that applied differential privacy (DP) mechanisms to perturb the gradients of each iteration and employed a privacy accountant to track the privacy loss during training. Thus, by limiting the impact of one single data on model parameters, the privacy can hardly be reverse-engineered. 

%As supposed to privacy-preserving training on sensitive data, an alternative approach is applying transfer learning to train a practicable model without exposing the senstive data.

%Existing literatures on privacy protection in deep learning mostly addressed three objective: the privacy of data used for learning a model, the privacy of the model, and privacy of the model's output. 
%Our key concern is the privacy of models accessible to adversaries and make sure that they cannot expose the training data with certain privacy guarantees. 

In practice, the aforementioned protections are prone to the worst case privacy guarantees assuming attackers have access to the internal model parameters, $i.e.$, very large noise can be injected at the cost of model utility degradation \cite{abadi2016deep}. Thus, the applicability of such mechanisms are questionable. A more promising alternative is to only expose a learned model obtained through transfer learning instead of the private model directly learned on the sensitive data. The training procedure with access to the private model is prudently privacy-bounded. For example, recent works \cite{papernot2016semi, papernot2018scalable} proposed a two-player (teacher-student) framework to train a student model on the differentially private aggregated outputs of an ensemble of teachers. Such privacy protection is achieved through the noisy voting by teachers but requires many teachers to compensate for the noise added to each query to ensure model utility. Although such a private transfer learning mitigates the exposure of private models associated with sensitive data, it still remains\textit{ a challenging task to balance between model utility and privacy when the training data is limited}. This is not a trivial problem and needs to resolve the following issues:
\begin{itemize}
    \item \textbf{Model performance bottleneck}. Differential privacy enforces a certain level of privacy budget given the composition theorem, inevitably restricting the number of training epochs that makes it hard to further improve model performance.
    \item \textbf{Availability of training resources}. For privacy purpose, it is desired to have only public data fed to all the networks. However, it is difficult for the student ($e.g.$, local reservoir of a hospital) to collect sufficient quality data for training ($e.g.$, lack of labels, limited quantity, $etc.$).
    \item \textbf{Overhead of noisy queries}. The standard approach to DP is to inject noise to the output while the noise scale is proportional to the sensitivity of the query function. A higher sensitivity inevitably results in excessive noise, completely masking the knowledge to be transferred. For example, PATE proposes to reduce the sensitivity of voting results by ensembling 250 teacher models \cite{papernot2016semi}. Such an approach brings significant overhead to computational resources and hence is not scalable to large-scale tasks. 
\end{itemize}

In this paper, we propose a novel three-player (teacher-student-discriminator) framework that combines the state-of-art knowledge transfer techniques with the advanced privacy-preserving mechanisms. 
In the framework, the target model is treated as the $student$ in the conventional \textit{teacher-student} learning paradigm and the generator in {generative adversarial networks} (GAN). The $student$ is not only trained by the $teacher$ through {Knowledge Distillation} (KD), but also adversarially trained with the $discriminator$ through GAN to generate similar outputs as the teacher. When with limited quality training data, which is often the case in practice, we enforce the discriminator to train the student to mimic the 'true' learned distribution of the teacher. The teacher in the framework is pre-trained on sensitive data and frozen during the training. Thus, unlike the two-way transfer learning as in \cite{wang2018kdgan}, the knowledge transfer in our framework is unidirectional, $i.e.$ only the teacher distills the knowledge to the student without privacy breach. Then, to enforce privacy guarantee during training, a privacy protection mechanism is introduced that provides insights into the vulnerabilities of the framework and applies DP to protect the private teacher. Finally, the proposed learning strategy achieves the joint utility and privacy optimization that transfers private knowledge from a protected teacher to a public student.

%A similar idea of three-player game has been discussed in \cite{wang2018kdgan}, where both the student and the teacher are adversarially trained against the discriminator to learn the true data distribution and then distill knowledge to each other.
%the target model is still treated as the $student$ as the conventional \textit{teacher-student} learning paradigm, the $teacher$ employs {Knowledge Distillation} (KD) with a softened distribution of the final output of the teacher network to teach information to the student. In addition, a $discriminator$ using {generative adversarial networks} (GAN) is introduced to force the student to generate similar outputs as teachers, eventually reaching equilibrium between the two. 
The proposed learning strategy for private knowledge transfer exhibits three important advantages: 
\begin{itemize}
    \item The student network that learns the distilled knowledge with the discriminator is better optimized than the conventional teacher-student paradigm;
    \item Faster convergence can be achieved even with limited training epochs, while the performance is not overly bounded by the number of training instances; 
    \item With the well-designed privacy-preserving mechanism, the framework is able to achieve excellent model utility and rigorous privacy guarantee.
\end{itemize} 
%With such a adversarial learning strategy, the propose framework can jointly optimize the distillation loss and the adversarial loss to reduce the number of training epochs. Then with an integrated privacy-preserving mechanism in the framework, we achieve \textit{a deep learning technique that transfers knowledge from the private teacher while balancing between good model utility and rigorous privacy guarantee}.
%$teacher$ and $student$ can reach equilibrium for good trade-off between privacy guarantee and model utility even with adequate learning epochs. (Need to rephrase this sentence... not accurate? ???adequate sounds like well tuned... your method should be more general and robust)} 
The contributions of this work can be briefly summarized as below:
\begin{itemize}
\item We propose a three-player (teacher-discriminator-student) framework to transfer private knowledge to the student. With the discriminator, the student can accurately and efficiently mimic the teacher even with limited quality training data, which enables excellent model utility and a strong privacy guarantee. 
\item We integrate a differential privacy mechanism into the learning procedure that allows student to be trained with a rigorous privacy budget, in which the privacy accountant provides a theoretical basis for trade-off between utility and privacy.
\item We evaluate the proposed framework on MNIST, SVHN and CIFAR-10 datasets. It is found that the proposed framework offers a good utility and privacy trade-off even with very few training epochs and unlabelled training instances. 
In addition,  Our students achieve accuracies of 98.52\%, 93.12\% and 84.79\% with DP bounds of (1.93, $10^{-5}$) for MNIST, (5.02, $10^{-6}$) for SVHN and (8.81, $10^{-6}$) for CIFAR-10, respectively.
Compared with the existing works, our framework consistently ensures a lower student accuracy loss ($w.r.t.$ the teacher). In particular, the students have only 0.89\%, 2.29\% and 5.16\% accuracy losses for the three datasets, while the reported student accuracy losses from other works are 1.21\%, 10.88\%, and 5.40\%, respectively \cite{papernot2016semi, wang2019private}.
\end{itemize}

%The framework can achieve a good trade-off between model utility and privacy guarantee using limited training data. 
%\redHL{need some transition, refer to \cite{xu2017training}} Due to the lack of labels of training data (in where???), the distillation loss is not sufficient to supervise the student network for which self cross entropy loss dominates the optimizing procedure. On the other hand, with a \textit{generative adversarial networks} (GAN), we further employ an adversarial learning strategy to force the student to generate similar outputs as teachers in which a discriminator can always make both of them reach equilibrium as long as adequate learning epochs. 
%This paper proposes to jointly optimize the distillation loss and the adversarial loss to reduce the number of training epochs required to converge. Then by applying the well-established differential privacy and the use of the RDP accountant, we achieve a deep learning techniques that transfer knowledge from the private teacher while keeps rigorous privacy guarantees. In summary, our contributions are as follows:

\section{BACKGROUND AND RELATED WORK}
Since the emergence of AlexNet, it is found that deeper neural networks require more training data to ensure convergence and robustness. In the healthcare, financial, or privacy-related domains, such training data is typically collected in a centralized manner and inevitably undergoes privacy breach risk if the trained models are exposed to the public. Thus, it is crucial to protect the privacy of training data as well as its use in deep neural networks. Recently various attacks further aggravate such privacy concerns in training and releasing deep learning models, as attackers can excessively analyze the model responses to recover the sensitive information. As a consequence, privacy protection in deep learning has been a concerning research area.

%Due to the rigorous law against data leakage, medical researchers attempt to generate realistic synthetic samples based on input real samples via techniques such as an autoencoder,GAN or a combination of the two\cite{choi2017generating}. However, such approach is not applicable to a wide range of scenarios since training datasets are so diverse that it is difficult to unify a technique to generate them while keeping high synthetic quality.  

\textbf{Differential privacy} (DP) is a widely-adopted approach to privacy protection in deep learning models. The theoretical study \cite{dwork2006calibrating} provides provable privacy guarantees for differential privacy that is achieved through adding noise to mask the output differences for the two different inputs. The very first proposal of deep learning with DP was presented in \cite{abadi2016deep}, in which the gradients in stochastic gradient descent (SGD) algorithm were perturbed and the privacy budgets were accordingly tracked using the \textit{moment accountant}. Its successful development has promoted several following studies \cite{mcmahan2017learning, yu2019differentially} on differentially private deep learning.

Instead of directly applying DP to model training, references \cite{papernot2016semi, papernot2018scalable} implemented a private transfer learning framework (PATE) that transferred the knowledge of an ensemble of teacher models to a student model. Intuitively, its privacy is guaranteed by training teachers on disjoint datasets and aggregating the outputs with noise. However, PATE requires a large number of teacher models to compensate for the noise injected to the individual query responses to ensure a desired trade-off between utility and privacy. In addition, the efficiency and effectiveness of PATE heavily relies on the correctness of voting results of the sample query as well as the appropriate noise added. This is actually hard in practice to select a voting label that is helpful to improve the training while does not reveal privacy when the student only has access to limited public (even unlabelled) data. Another representative work is \cite{wang2019private}, which introduced a private model compression framework with the conventional transfer learning and supervised learning techniques. However, the efficiency and applicability of the work \cite{wang2019private} may be significantly impacted when the student only can use limited quality data.
%The efficiency and effectiveness of PATE is then very limited when the student/learner only has access to a limited amount of public unlabelled data, which however is often the case in practice.
%\redHL{YOU can talk more about the difference and issue here for PATE}

\textbf{Privacy accountant} is an indispensable part to DP, which can track the accumulated privacy loss and enforce the applicable privacy policy \cite{abadi2016deep}. It has been noted that the privacy loss radically comes from the number of queries responded by the private teacher, $i.e.$, the number of iterations during student training. A formal definition of differential privacy \cite{dwork2011differential} is given as below : 
\begin{Def}
A randomized mechanism $\mathcal{M}$ with domain $\mathcal{D}$ and range $\mathcal{R}$ satisfies ($\varepsilon,\delta$)-differential privacy if for any two adjacent inputs $D$, $D'\in\mathcal{D}$ and for any subset of outputs $\mathcal{S}\subseteq\mathcal{R}$ it holds that:
\begin{equation}
\begin{aligned}
Pr[\mathcal{M}(D)\in\mathcal{S}]\leq{e}^{\varepsilon}\cdot{Pr}[\mathcal{M}(D'\in\mathcal{S}]+\delta.
\label{def:dp}
\end{aligned}
\end{equation}
\end{Def} 
In Eq.~\eqref{def:dp}, the parameter $\epsilon$ is an upper bound for privacy loss, and parameter $\delta$ is a failure probability for this privacy guarantee. A smaller privacy bound $\epsilon$ enforces a stronger privacy guarantee but inevitably incurs more significant accuracy loss. Based on the composition theorem of DP, \cite{abadi2016deep} proposed the concept of \textit{moment accountant} to quantitatively capture the privacy budget. The extended work in \cite{mcmahan2018general, mironov2019r} further proposed the application of R{\'e}nyi differential privacy (RDP) accountant by analyzing R{\'e}nyi divergence to enable tighter bounds of privacy loss. In this paper, we employ \textit{RDP accountant} to measure the privacy budget of the proposed training procedure, and balance the trade-off between utility and privacy.

It is noted that direct applications of DP to the exposed models may incur too much utility loss. 
Other than transfer learning via an ensemble of teachers, \textbf{Knowledge distillation} (KD) is considered as an effective alternative to transfer knowledge from a private teacher to a student network to protect the privacy of training data. In the teacher-student paradigm, the teacher teaches the student through a softened distribution of its output, $a.k.a.$, dark knowledge \cite{xu2017training}. It is easier to mimic the teacher than directly learning the target function, as the output distribution from the teacher embodies additional information beyond the ground-truth distribution. A recent study in \cite{lopez2015unifying} provided theoretical insights that an intelligent teacher can transfer helpful privileged information only available at the teacher's training stage. Thus, the performance of the published student model is actually associated with the distilled knowledge from the teacher network trained on private data. Such findings motivate us to investigate how to design knowledge transfer to balance between utility and privacy.

\textbf{Generative adversarial network} (GAN) is another alternative to enable the model to learn the true data distribution \cite{isola2017image,yu2017seqgan,zhang2017adversarial}. GAN may employ a generator to synthesize student features and a discriminator to distinguish student outputs from teacher outputs. Recently GAN has been successfully applied to generate discrete data ($e.g.$, sequence, text) \cite{yu2017seqgan, zhang2017adversarial}. This is consistent with our purpose of generating/learning a discrete classification distribution for the student.

%Due to the fact that discrete data makes it difficult to pass gradients from a discriminator backward to update a generator, we employ the approach of Gumbel-Max trick same with \cite{wang2018kdgan} to relax the discrete distributions learned by the generator into continuous distributions.  
Thus, in this paper, we explore the idea of integrating KD and GAN, where \textit{a discriminator trains the student to learn the distribution over the pseudo labels created by the teacher while the teacher distills dark knowledge to the student.} An earlier work of \cite{wang2018kdgan} also proposed to combine KD with GAN to help speed up the training procedure, where both the student and the teacher are trained against the discriminator to learn the true distribution and then distill knowledge to each other. Such two-way transfer learning inevitably undergoes the risk of privacy breach and hence cannot be simply applied to the private knowledge transfer.

%It is observed that the optimization problem of learning to mimic the teacher network turns out to be easier than learning the target function directly \cite{romero2014fitnets}, it is due to that teachers' classification probabilities and/or features representation conveys addition information beyond the supervised learning of ground-truth distributions.
 %Here we address distilling a protected network trained on sensitive and/or private data for publishing of a student network accessible to all, in which the student's performance is assured by effective knowledge distillation.

\begin{figure*}[htp]
    \centerline{\includegraphics[width=18cm]{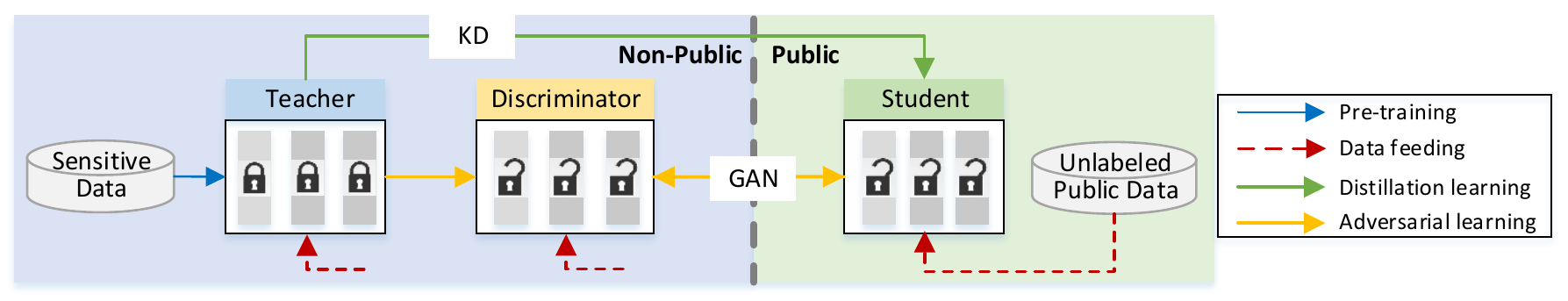}}
    \caption{Overview of the proposed framework and its data flow.} 
    \label{fig_overview}
\end{figure*}

\section{OVERVIEW}

This paper proposes a private knowledge transfer strategy, where the private teacher trained on sensitive data is not publicly accessible but can be used to teach a student model to be released. The target student is not only taught by the distilled knowledge from the teacher, but also trained against a discriminator to mimic the behavior of the teacher. There are two critical questions to be answered to implement the aforementioned strategy: (1) How to combine KD with GAN to train the target student model? (2) How the teacher response is privacy-guaranteed during the training procedure? 

An overview of the proposed framework to implement such a strategy is demonstrated in Figure~\ref{fig_overview}. The framework involves three players (teacher-student-discriminator) and two domains (non-public and public). 
 %\textit{knowledge transfer paths} (KTP) 
There are two \emph{knowledge transfer paths} for the target student. KD acts as a unidirectional path from teacher to student, and GAN is a two-way path between discriminator and student:
\begin{itemize}
    \item\textbf{Distillation learning by KD} {(detailed in Sec. 4.1):} For a given input instance, we adopt the Kullback Leibler (KL) divergence as the distillation loss to measure the distance between two categorical distributions of teacher and student, $i.e.$, class probabilities from teacher and student. The effectiveness of distillation stems from the additional supervision and regularization of higher entropy soft targets.
    \item \textbf{Adversarial learning by GAN} {(detailed in Sec. 4.2):} Student (generator) and discriminator play a min-max game between one another. The discriminator tries to distinguish the student's output from the teacher's, while the student tries to generate similar output as the teacher that makes the discriminator can no longer differentiate. Both student and discriminator are adversarially trained epoch by epoch until the equilibrium. 
\end{itemize}

Due to the lack of high quality training data for the student, it is typically challenging for the student to use the cross-entropy error as the objective function, which makes the training without an effective supervision. 
However, the combination of knowledge distillation and adversarial learning results in more effective optimization that resolves the issue. 
As discussed in \cite{wang2018kdgan}, a weighted sum of the distillation loss (from the teacher) and the adversarial loss (from the discriminator) may reduce the gradient variance, thereby accelerating the student training convergence with fewer training epochs. Motivated by that, we propose a joint optimization of distillation and adversarial losses (as detailed in Sec. 4.3) to cover the \textit{first question}, which can enforce the student to accurately mimic the teacher and ensure GAN to quickly reach the equilibrium.

To protect the privacy of sensitive data, as shown in Figure~\ref{fig_overview}, we keep the pre-trained teacher model (on the sensitive data) inaccessible to the public (or attackers), $i.e.$, either internal model parameters or outputs of the model are not available to the public. The only accessible part in the framework is the student model, which can take nonsensitive public data. The connection between non-public and public domains are the two knowledge transfer paths without any direct access to private data. Since the student itself cannot access the private data, the remaining privacy protection problem for the framework is how to limit privacy loss for querying the private teacher during the student training. 

Intuitively, the excessive memorization of the private teacher may expose the private and sensitive data under attacks. Thus, we propose to integrate a differential privacy (DP) mechanism into the training procedure for privacy guarantee while ensuring model utility (as detailed in Sec. 4.1), which answers the \textit{second question} for the strategy implementation. The mechanism injects the Gaussian noise to each query to the teacher, and then track the bound of privacy loss based on the Composition Theorem of DP \cite{dwork2011differential}. In other words, we sanitize the teacher's distillation loss through clipping the batch loss with the global norm bound and perturbing it with appropriate Gaussian noise. Since the discriminator in the framework is not accessible to attackers after the training, the discriminator itself and its adversarial training procedure does not incur additional privacy violations. Finally, the mechanism employs \textit{RDP accountant} to keep track of privacy loss during training, enabling a theoretical basis for trade-off between utility and privacy.

\begin{figure}[!htb]
    \centerline{\includegraphics[width=10cm]{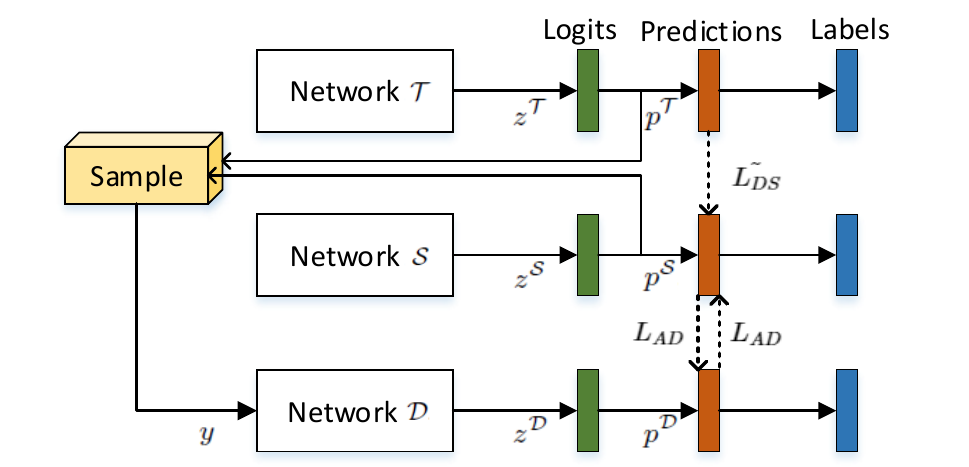}}
    \caption{Private knowledge transfer strategy: the student network is trained with adversarial and distillation losses; the discriminator network is trained to distinguish the sampled probability distributions of teacher and student. } 
    \label{fig_learning}
\end{figure}

\section{Method}

In this section, we formulate the proposed learning strategy with a cohort of the three networks (as plotted in Figure~\ref{fig_learning}). Given a private dataset, we can always pre-train the teacher network $\mathcal{T}$ using the cross-entropy error as the objective function. On the other hand, with {a nonsensitive public dataset $X$}, the student $\mathcal{S}$ is trained by not only the teacher $\mathcal{T}$ to minimize the perturbed distillation loss $\tilde{L_{DS}}$, but also the discriminator $\mathcal{D}$ to minimize the adversarial loss $L_{AD}$. In the following, We first discuss the knowledge transfer path of KD and its training loss. Then, we introduce the proposed privacy protection mechanism. After that we describe the knowledge transfer path of GAN and the adversarial learning loss. Finally, we present the joint learning procedure for the framework.

\subsection{Knowledge Transfer Path with KD}

\subsubsection{Knowledge distillation}
Given a sample set of $N$ samples $X = \{x_i\}_{i=1}^N$ from $M$ classes, we denote the corresponding label set as $Y = \{y_i\}_{i=1}^N$ with $y_i = \{1,2,...,M\}$. The categorical distributions of the two networks $\mathcal{T}$ and $\mathcal{S}$ are probability values of $M$ classes, denoted by $p^\mathcal{T}$ and $p^\mathcal{S}$, respectively. For a sample $x_i$, the two probabilities are:
\begin{eqnarray}
    p^\mathcal{T}(x_i) = softmax(z_i^\mathcal{T}) \quad \textrm{and} \quad
    p^\mathcal{S}(x_i) = softmax(z_i^\mathcal{S})
\label{equ_cp}
\end{eqnarray}
where the logits $z^\mathcal{T}$ and $z^\mathcal{S}$ are the outputs of the last fully connected layer of the two networks $\mathcal{T}$ and $\mathcal{S}$, respectively. 

Kullback Leibler (KL) Divergence is a measure of how one probability distribution differs from another. Here we adopt KL divergence as the distillation loss to distill knowledge from network $T$ to network $S$. Then we have the following loss function:
\begin{equation}
    L_{DS} = 
    \sum_{i=1}^{N} p_{\boldsymbol{\tau}}^\mathcal{T}(x_i){\rm log}(\frac{p_{\boldsymbol{\tau}}^\mathcal{S}(x_i)}{p_{\boldsymbol{\tau}}^\mathcal{T}(x_i)})
\label{equ_ds}
\end{equation}
where the probability distribution $p_{\boldsymbol{\tau}}(x_i)=softmax(z_i/{\boldsymbol{\tau}})$ is the softmax temperature function with ${\boldsymbol{\tau}}$ as the temperature parameter. The temperature parameter ${\boldsymbol{\tau}}>0$ controls how much we want to soften or smooth the class probability predictions from $p^\mathcal{T}$. A higher temperature indicates a softer probability distribution generated by the teacher network with privileged knowledge of the differences among the classes.

\subsubsection{Differential privacy protection}
The proposed privacy protection is built upon a general approach in \cite{mcmahan2018general}, which can allocate privacy budget to each step and compute the total privacy cost over iterations.

Given a probability sample $q$, clipping threshold $C$, noise multiplier $m$,  in \cite{mcmahan2018general}, the standard procedure of adding Gaussian noise ($i.e.$, applying DP) to a vector to be protected is:
\begin{enumerate}[\indent 1.]
    \item Select a subset of records (or samples) $R_i$, $i\subseteq[1,...,N]$, with probability $q$ to choose each record. The result of each query for the record is a vector $v^i\in{R^D}$;
    \item Clip each $v^i$ by threshold $C$ and $L_2$ norm: $\pi_{C}(v^i)=v^i{\cdot}{\rm min}(1,C/||v^i||_2)$;
    \item The sum with noise (or DP) added is then computed by: $\tilde{v}=\frac{1}{qN}(\sum\pi_{C}(v^i)+\mathcal{N}(0;\sigma^2I))$.
\end {enumerate}
where $\sigma=m\cdot{C}$ and can be intuitively understood as the scale of the injected noise.

Unlike the general approach in \cite{mcmahan2018general} that applys DP to the gradients during training, the primary privacy concerns of the proposed strategy (as briefly discussed in Sec. 3) is only associated with the private teacher network, in particular, the distillation loss to be queried. Thus, in the proposed framework, we do not need to protect the gradient vector as \cite{mcmahan2018general}. Instead, we can choose a batch of samples, then protect a vector of batch distillation loss that is queried from the teacher. We can use the following to compute the batch distillation loss: 
 \begin{equation}
     v^i = p_{\boldsymbol{\tau}}^\mathcal{T}(x_i){\rm log}(\frac{{p_{\boldsymbol{\tau}}^\mathcal{S}(x_i)}}{{p_{\boldsymbol{\tau}}^\mathcal{T}(x_i)}})
     \label{equ_ds}
 \end{equation} 
To protect the vector, we can then inject Gaussian noise with mean of 0 and standard deviation of $m\cdot{C}$ to the norm-bounded batch distillation loss. As a result, the batch of differentially-private distillation loss can be defined as:
\begin{equation}
    \tilde{L_{DS}} =\frac{1}{qN}  (L_{DS}   / {\rm max}(1,\frac{||L_{DS}||_2}{C} ) + \mathcal{N}(0; \sigma^2I))
\label{equ_dp_ds}
\end{equation}
With such a differentially-private loss $\tilde{L_{DS}}$ for each batch of samples, we can bound the privacy cost of each query to the teacher. In addition, it is necessary to keep track of the privacy cost during the entire training procedure and then derive the final privacy budget for a desired model utility. Such a relationship acts as the theoretical basis for our trade-off between utility and privacy. 
We employ the concept of R{\'e}nyi divergence and R{\'e}nyi differential privacy (RDP) accountant \cite{mironov2017renyi, mironov2019r} as the measure. Per the definitions of DP and RDP, the privacy loss can be considered as a random variable dependent on the injected random noise. 
RDP accountant then generalizes the pure $\epsilon$-differential privacy and achieves a more compact composition theorem than the standard ($\epsilon$, $\delta$)-DP \cite{mironov2017renyi, mironov2019r}. In particular, RDP ensures that a randomized mechanism can be bounded by a smaller $\epsilon$ through R{\'e}nyi divergence of two adjacent inputs, and then tracks privacy bounds on the moments of the privacy loss. In the proposed DP mechanism, RDP is applied to track the bound on the batch distillation loss with sampled Gaussian noise.
%Here we recall the definitions of R{\'e}nyi divergence, R{\'e}nyi differential privacy. 

%\bluHL{The proposed DP protection mechanism is essential to make a trade-off between model utility and privacy bound for a given task. }

\subsection{Knowledge Transfer Path with GAN}
With the pre-trained teacher network $\mathcal{T}$, the student network $\mathcal{S}$ can be trained adversarially against the discriminator network $\mathcal{D}$. The student and the discriminator play a min-max game. While the network $\mathcal{S}$ attempts to generate a probability $p^\mathcal{S}$ mimicking the distribution of the teacher network $p^\mathcal{T}$, the discriminator model tries to distinguish the 'true' label predicted by $\mathcal{T}$ from the pseudo label by $\mathcal{S}$. We then can define the objective function $L_{AD}$ for the min-max game:
\begin{eqnarray}
    \mathop{\rm min}\limits_s\mathop{\rm max}\limits_dL_{AD} = 
    E_{y\sim{p^\mathcal{T}}}[{\rm log}\,p^\mathcal{D}(y) + \nonumber\\
    E_{y\sim{p^\mathcal{S}}}[{\rm log}\,(1-p^\mathcal{D}(y)]
\label{equ_gan}
\end{eqnarray}
where $y\sim{p^\mathcal{T}}$ and $y\sim{p^\mathcal{S}}$ are the continuous samples generated from the discrete probability distributions of $p^\mathcal{T}$ and $p^\mathcal{S}$, respectively; $p^\mathcal{D}(y)$ is the probability generated by the discriminator network for a label $y$. 

In the proposed framework, the network $\mathcal{D}$ gets updated by maximizing the objective function $L_{AD}$ in Eq.~\eqref{equ_gan}, while $\mathcal{S}$ attempts minimizing $L_{AD}$, thereby making $\mathcal{D}$ unable to differentiate whether a given label is predicted by $S$ or not. Such a min-max game updates $\mathcal{S}$ and $\mathcal{D}$ alternatively until the equilibrium is reached, $i.e.$, $\mathcal{S}$ learns the distribution of $\mathcal{T}$ given the discrimination of $\mathcal{D}$. As shown in Figure~\ref{fig_learning}, the network $D$ cannot directly take the discrete probabilities from $\mathcal{T}$ and $\mathcal{S}$, which may inevitably result in high variances in the gradients. To reduce the variances for $\mathcal{D}$, we use the Gumbel-Max trick \cite{maddison2014sampling} to re-parameterize the generation of the discrete samples to a continuous space. Then, we can conduct sampling approximation to obtain the continuous samples $y$ and formulate $L_{AD}$ with lower-variance gradients.

\subsection{Joint Learning Procedure}
Based on the two knowledge transfer paths and the proposed privacy protection mechanism, we incorporate the differentially private distillation loss in Eq.~\eqref{equ_dp_ds} and adversarial loss in Eq.~\eqref{equ_gan} into the final objective function for the target student as below:
\begin{eqnarray}
   L = \alpha\tilde{L_{DS}} + (1 - \alpha)L_{AD}
\label{equ_total}
\end{eqnarray}
where $\alpha$ is a distillation weight set between 0 and 1. %\redHL{We set it as 1 in the experimental part for ablation study.}
We achieve the joint utility and privacy optimization with the proposed privacy protection mechanism that transfers knowledge from a protected private teacher to a public student. 
The overall learning procedure is summarized in Algorithm~\ref{algorithm}. 

Our learning strategy has its unique advantages: when combined with the adversarial learning, the joint optimization of distillation loss and adversarial loss can greatly save training epochs, which results in a stronger privacy guarantee. In addition, with RDP accountant, by searching the optimal values for the hyper-parameters ($i.e.$, $q$, $C$, $\sigma$) and the distillation weight $\alpha$, we can ensure an optimal model utility with a tight privacy bound. 

\begin{algorithm}
	\caption{Privacy-Preserving Learning Procedure.}
    \label{algorithm}
    \KwIn{a pre-trained network $\mathcal{T}$, public training samples $N$, training epochs $T$, training epochs for the discriminator $T_\mathcal{D}$, training epochs for the student $T_\mathcal{S}$, batch size $B$, clipping threshold $C$, the noise multiplier $m$.}
    \KwOut{a student network $\mathcal{S}$}  
    \For{$t=0$ to $T-1$}
    {
        \For{$i=0$ to $(T_D-1)*(N/B)$}
        {
            Sample a batch $x$ of size $B$;\\
            Sample $y$ from discrete probabilities $p^\mathcal{T}$ and $p^\mathcal{S}$;\\
            Compute adversarial loss $L_{AD}$;\\
            Update $\mathcal{D}$ by ascending along its gradients of $L_{AD}$ ;
        }
        \For{$j=0$ to $(T_S-1)*(N/B)$}
        {	
        	Sample a batch $x$ of size $B$;\\
            Compute distillation loss $L_{DS}$;\\
            Apply differential privacy mechanism: $\tilde{L_{DS}} \gets L_{DS}  / {\rm max}(1,\frac{||L_{DS}||_2}{C} ) + N(0; \sigma^2I)$;\\
            Sample $y$ from discrete probabilities $p^\mathcal{T}$ and $p^\mathcal{S}$;\\
            Compute adversarial loss $L_{AD}$;\\
            Compute weighted sum $L$;\\
            Update $\mathcal{S}$ by desending along its gradients of $L$;
        }
    }
\end{algorithm}

\section{EXPERIMENTAL RESULTS}

The proposed framework is applicable to a wide range of multi-label learning tasks, where the students can learn from the teachers owning sensitive private data. Note that the privacy budget of a complete training procedure greatly depends on the noise injected to each training step and the number of training epochs. To prove the generality, we here employ three widely-adopted datasets, MNIST, SVNH, CIFAR, to evaluate the proposed framework. We compare our results with the state-of-art existing works in \cite{papernot2016semi, wang2019private} using the reported data of \cite{papernot2016semi} on MNIST and SVHN, and the reported data of \cite{wang2019private} on CIFAR-10, respectively. %In this section, we demonstrate the better perforamnce and faster convergence of the proposed framework offers a better trade-off between utility and privacy than previous work.

\subsection{Experiment Setup}
Here we briefly describe our experiment setup. We implement our learning strategy based on Tensorflow.  The experiments are based on MNIST, SVHN and CIFAR-10 classification tasks. We first pre-train a teacher model on the entire dataset (with separate training and testing), and treat it as the private teacher. Then we randomly select the public data  from the training dataset and assume that those data are unlabelled, which is used to train the student model through the proposed learning strategy in Sec. 4. 

Three datasets of MNIST, SVNH and CIFAR-10 are used in our experiments with the following details:

\textbf{MNIST.} The MNIST dataset  \cite{lecun1998gradient} has 60000 grayscale images (50000 for training and 10000 for testing) with 10 different label classes. Teacher, student and discriminator are implemented using a LeNet, an MLP and a LeNet. When trained on the entire dataset, the teacher model has a 99.40\% test accuracy. We vary the number of unlabelled training instances in [100, 10000] to train the student.

\textbf{SVHN.} The SVHN dataset \cite{netzer2011reading} consists of $32\times32$ colored digit images, each digit representing one class. The training and testing sets contain 604388 and 26032 images, respectively. Teacher, student and discriminator are implemented using a ResNet, a LeNet and a LeNet. The teacher has a 95.30\% test accuracy after training. The number of unlabelled training instances to train the student is varied in [500, 50000].

\textbf{CIFAR.} The CIFAR-10 dataset \cite{krizhevsky2009learning} contains colored natural images with a size of $32\times32$, with 10 classes. The training and testing sets contain 50000 and 10000 images, respectively. The three networks and training setup are the same as the SVHN dataset. The teacher can reach a 89.4\% accuracy after training.

\begin{figure*}[htp]
	 \caption{Training curves (blue) of teachers on the private dataset, and training curves (red) of students on a public subset with 10000 training instances for the three datasets: (a) MNIST; (b) SVHN; (c) CIFAR-10.}
	\label{fig:trainingcurves}
    \begin{minipage}[t]{0.33\linewidth}
    \centering
    \includegraphics[width=6cm]{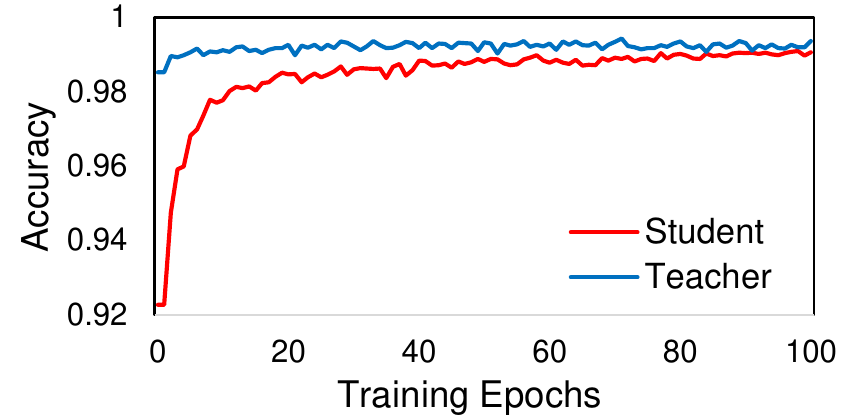}
    \caption*{(a) Training curves for MNIST}
    \label{fig:curves:a}
    \end{minipage}
    \begin{minipage}[t]{0.33\linewidth}
    \centering
    \includegraphics[width=6cm]{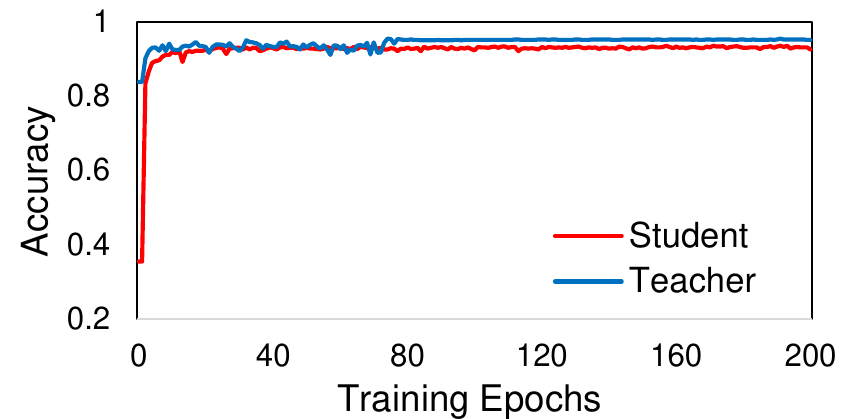}
    \caption*{(b) Training curves for SVHN.}
    \label{fig:curves:b}
    \end{minipage}
    \begin{minipage}[t]{0.33\linewidth}
    \centering
    \includegraphics[width=6cm]{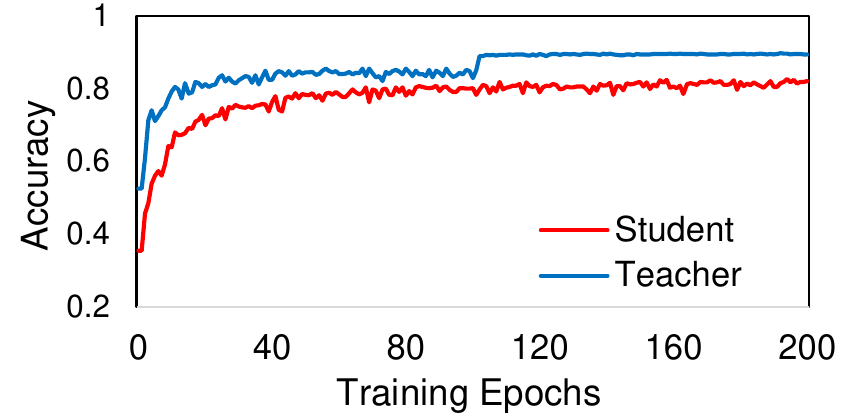}
    \caption*{(c) Training curves for CIFAR-10.}
    \label{fig:curves:c}
    \end{minipage}  
\end{figure*}

\begin{table*}[tbp]
\newcommand{\tabincell}[2]{\begin{tabular}{@{}#1@{}}#2\end{tabular}}\vspace{-0.3cm}
    \caption{Comparison on average accuracy over 20 runs between the proposed joint learning strategy (combining KD and GAN) and the KD-only optimization for different number of training instances (denoted as $n$) for students.}\vspace{-0.3cm}
    \label{table:train:size}
    \begin{center}
        \begin{tabular*}{17cm}{@{\extracolsep{\fill}}cccccccccc}
            \hline
            \multirow{2}*{Method} & \multicolumn{3}{c}{MNIST} & \multicolumn{3}{c}{SVHN} & \multicolumn{3}{c}{CIFAR} \\ 
            \cline{2-4} \cline{5-7} \cline{8-10}
 			& n=100 & n=1000 & n=10000 & n=500 & n=5000 & n=50000  & n=500 & n=5000 & n=50000 \\ \hline
 KD-only (\%) & 66.61 & 93.57 & 98.38& 33.24 & 86.28 & 92.49 & 19.71 & 64.59 & 81.18 \\
\hline
Joint (\%) & 67.99 & 96.46 & 98.52& 56.23 & 87.74 & 96.62 & 46.45 & 66.9 & 80.06 \\
 \hline
        \end{tabular*}
    \end{center}\vspace{-0.3cm}
\end{table*}

\begin{figure}\vspace{-0.3cm}
    \centerline{\includegraphics[width=8cm]{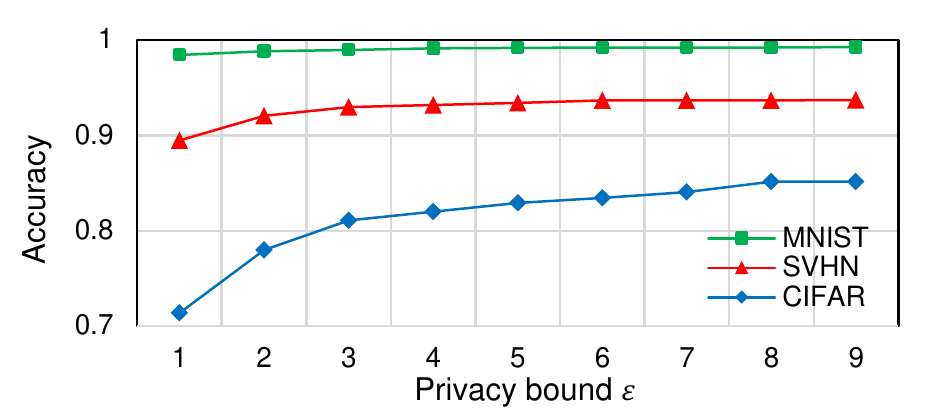}}
    \caption{Student model accuracy $v.s.$ privacy bound $\epsilon$ for the three datasets.} 
    \label{fig:acc:bound}\vspace{-0.3cm}
\end{figure}

\subsection{Results and Analyses}

The typical application of knowledge distillation is to transfer from a powerful and large network or an ensemble of networks to a small network (also known as deep model compression), which can reduce the network complexity and capacity to improve the deployability of the deep models \cite{wang2019private,wang2018kdgan,xu2017training}. In this paper, instead of focusing on the model compression performance, \textbf{the goal is to achieve a good student accuracy under a tight privacy bound, $i.e.$, optimal trade-off between utility and privacy.} We conduct experiments with different number of unlabelled data to explore how utility and privacy budget vary against the number of training instances and training epochs. This is aligned with the aforementioned motivation of this paper and the practical demands from mobile/edge scenarios.

In the following, we thoroughly study the impacts from training size, epochs, and privacy bounds on the framework performance. For the other hyper-parameters in Algorithm \ref{algorithm} ($e.g.$, batch size, distillation weight, clipping threshold, noise multiplier, $etc.$), the optimal values within the range are searched and pre-determined. It is noted that, even with such practical constraints of tight privacy bound, fewer training epochs and limited public data, the proposed framework still can exhibit higher model utility than prior works, with different network typologies employed to teachers and students ($e.g.$, a MLP student and a LeNet teacher for MNIST). This indicates the overall superiority of the framework and its applicability to real-world mobile/edge scenarios with limited resources. 

\textbf{Training speed.} We first investigate the training speed of the framework on the aforementioned three datasets. The learning curves of both teachers and students are plotted in Figure~\ref{fig:trainingcurves}, where the students are trained with a subset of 10000 public unlabelled training instances. It is intuitively understandable that, in knowledge distillation, the student model accuracy learned from the teachers is sub-optimal to the teacher. Thus, we observe in the figure that the student models for MNIST, SVHN, CIFAR can quickly reduce the accuracy loss to 0.89\% after 20 training epochs, 2.34\% after 20 training epochs, and 13.69\% after 30 training epochs, respectively. As discussed in Sec. 4, Figure~\ref{fig:trainingcurves} also validates that our student models can reach the convergence with a small number of training epochs (20-30). Thus, the proposed learning strategy that combines KD with GAN greatly \textbf{speeds up the training procedure by reducing the number of training epochs for convergence}. As a result, the fewer accesses to the private teacher induces a smaller privacy bound and cost to facilitate a meaningful utility .

\textbf{Training size.} The proposed strategy combines the advantages of both KD and GAN: 1) KD requires a small number of training instances to distill knowledge; 2) GAN enforces the student to learn the true distribution in the min-max game. We compare the proposed joint optimization with the KD-only optimization to show the efficiency of the proposal, the results of which are summarized in Table~\ref{table:train:size}. We vary the size of training sets on MNIST, SVHN and CIFAR-10 and compute the average accuracy over 20 runs. It is observed that the joint optimization of KD and GAN consistently outperforms the KD-only optimization, especially when only a small number of training images are available. For example, there are a 22.99\% accuracy improvement for SVHN and a 26.74\% accuracy improvement for CIFAR-10 when with 500 training instances. Thus, the joint optimization of KD and GAN \textbf{ not only reduces the required number of training instances but no longer need to purposely select the training instances }to achieve equally effective student model. This is especially beneficial to healthcare and finance applications where local agents (students) typically have very few data for training.
%from the teacher to the student but its supervison alone is not sufficient to converge; 
%2) GAN is well suited for the student to adversarially learn the true distribution at the equilibrium but it maybe take excessively long time for such a two-player game. 
 
%This observation is consistent with our above discusstions that the combination of distillation loss and adversarial loss can achieve a desired accuracy when a smaller training instances are available and even fewer training epochs are afford. This can be explained 

\textbf{Utility $v.s.$ privacy.} To study the trade-off between utility ($i.e.$, student model accuracy or accuracy loss $w.r.t.$ the teacher) and privacy ($i.e.$, differential privacy bound $\epsilon$), we track the values of $\epsilon$ for a given failure probability $\delta$ in the proposed differential privacy mechanism through the RDP accountant. We set the experiments with training size of 10000, batch size $B$ of 50, noise multiplier $m$ of 1.1. In Figure~\ref{fig:acc:bound}, we report the values of the $\epsilon$-differential privacy bounds ($\delta$ is $10^{-5}$ on MNIST, $10^{-6}$ on SVHN and CIFAR-10) for student training and testing. It is found that with a smaller $\epsilon$ guarantee, our MNIST student can still achieve a high accuracy ($i.e.$ ($1.93$, $10^{-5}$)-differential privacy for a $98.52\%$ accuracy). For SVHN and CIFAR-10, student accuracies generally increase with growing privacy bound but quickly converge after certain privacy bound thresholds ($i.e.$, SVHN student's accuracy stays almost the same for privacy bound $\epsilon > 3.5$, and CIFAR-10 student's accuracy for a bound $\epsilon > 8.5$). 
Thus, the proposed framework is capable to \textbf{effectively make trade-off between utility and privacy, allowing the student to achieve an optimal accuracy within a small privacy bound}.

\subsection{Comparisons and Discussions}

We compare the proposed framework with two state-of-art related works, PATE \cite{papernot2016semi} and RONA \cite{wang2019private}. As is discussed in the previous sections, the major differences among the three methods are: for PATE, the student receives knowledge from an ensemble of teachers, whereas for RONA, the student is trained with cross entropy loss supervised from labelled data in addition of distillation loss. As in Sec. 5.2, the proposed framework has already been proved to achieve fast training speed with fewer training instances (that can be unlabelled), exhibiting superiority for privacy-concerned applications in mobile/edge scenarios. Thus, the focus of comparison here is placed upon the utility and privacy trade-off. 

Table~\ref{table:comparison } compares the performance of the three methods, which reports the student accuracies and the accuracy losses $w.r.t.$ teachers for a given privacy bound. For fair comparison, the reported data of PATE and RONA in Table~\ref{table:comparison } are directly taken from the papers \cite{papernot2016semi,wang2019private} to prevent any re-implementation issues. Note that RONA requires all the public data are labelled \cite{wang2019private}, which is different from the setup of PATE and the proposed framework. However, since PATE does not have the results for CIFAR-10, we compare with RONA on CIFAR-10 to demonstrate the scalability of the proposed framework (even under unfavorable conditions). It is observed that the proposed framework can offer better trade-offs between privacy ( privacy bound $\epsilon$) and utility (student model accuracy) than the other two methods for all the datasets.
Moreover, with the relaxation of privacy bound $\epsilon$, the proposed framework can achieve more significant accuracy improvement than the other works. For examples, when $\epsilon$ is increased from 2 to 8 for MNIST, the accuracy loss of our student is reduced from 0.89\% to 0.20\%, while the change for PATE is very small, from 1.21\% to 1.11\%, indicating an 82\% and 26\% \textit{accuracy loss improvements} (defined as the difference between the accuracy losses for two methods over the loss for the reference method) for the two privacy bounds.

Such performance difference is due to the fact that the the accuracy improvement of PATE requires a massive number of queries, while its privacy budget sharply grows with increasing queries \cite{papernot2016semi}.
%the noisy voting among all the teachers, in which the accuracy improvement requires a massive number of queries. 
On the other hand, the proposed mechanism applies Gaussian noise with random sampling to the loss function for DP guarantee and is shown with good scalability, as the privacy cost of single query is quadratically reduced with the sampling rate.
%automatically reduce the number of accesses and privacy breach risk from each query.
Such scalability limitation of PATE can be amplified when with more complex dataset. For example, for SVHN, with an $\epsilon$ of 5.04, the student model from PATE has an accuracy of 82.70\% $w.r.t.$ a 92.80\% teacher accuracy, resulting in significant accuracy loss. Unlike PATE, with a similar $\epsilon$ of 5.02, our framework can achieve 93.12\% student model accuracy for a 95.30\% teacher accuracy ($i.e.$, 79\% accuracy loss improvement from PATE), making it feasible to achieve a desired accuracy and privacy trade-off. 

For the comparison with RONA, with $\epsilon$ bounds of 4.21 and 8.81, our student can achieve accuracies of 81.76\% and 84.79\%, which is slightly better than the performance of RONA\cite{wang2019private} ($i.e.$, 5\% accuracy loss improvement from RONA). However, as is discussed earlier, RONA relies on both transfer and self learning where all the public samples must be labelled. Such scenarios can be limited in practice. Unlike that, our framework can achieve good performance even with training data unlabelled, which is a more challenging but practical scenario, indicating a broader applicability.

Note that the proposed framework can be used as a baseline backbone for private knowledge transfer even with limited quality public data, which is well aligned with the goal of this paper to develop the transfer framework instead of extensive optimizations. In other words, many other optimization techniques, such as well-designed sample selection for querying, carefully-analyzed sensitivity, can be seamlessly integrated into this framework to further improve model utility or trade-off between utility and privacy. For example, the selective aggregation mechanism in improved PATE \cite{papernot2018scalable} can reduce the added noise and achieve an accuracy loss of 0.71\% for a (1.97, $10^{-5}$) bound on MNIST. Such optimization techniques are orthogonal to the proposed privacy knowledge transfer framework and hence can be always employed at the cost of complexity, which is not the focus of the paper, but can be explored in the future.

%Finally, we also compare the proposed framework with improved PATE in \cite{papernot2018scalable}, which demands a more selective aggregation mechanism to select the appropriate samples so as to reduce the added noise and achieve a tighter privacy budget. At the cost of such a selective mechanism, the improved PATE can achieve an accuracy of 98.5\% with a loss of 0.71\% for a (1,97, $10^{-5}$) bound on MNIST, which is slightly better than the proposed framework. While improved PATE may meet with similar scalability issue as PATE \cite{papernot2018scalable,papernot2016semi}, it is also noted that such a selection mechanism is orthogonal to the privacy knowledge transfer and hence can be always integrated into our framework at the cost of complexity. Since the focus of this work is on private knowledge transfer with limited quality instances, such selection and querying alternatives will be explored in the future work.  %There are several promising approaches: well-designed sample selection for querying, carefully-analyzed sensitivity and norm bound parameter, more accurate privacy account, $etc.$ %Since the focus of this paper is to discuss how the privacy benefits from the combination of KD and GAN, we could leave the detailed privacy mechanism exploration to the future work.  

 \begin{table}
\newcommand{\tabincell}[2]{\begin{tabular}{@{}#1@{}}#2\end{tabular}}\vspace{-0.1cm}
    \caption{Utility and privacy trade-off comparisons among the proposed framework (Proposed), PATE \cite{papernot2016semi} and RONA \cite{wang2019private}}.\vspace{-0.3cm}
    \label{table:comparison }
    \begin{center}
        \begin{tabular}{|c|c|c|c|c|c|}
            \hline
            \multirow{2}*{Dataset} & \multirow{2}*{Framework} & Privacy & \multicolumn{2}{c|}{Accuracy} & Accuracy \\ 
            \cline{4-5} {} & {} & Bound $\epsilon$ & Student & Teacher & loss  \\ 
            \hline\hline
            \multirow{4}*{MNIST} & \multirow{2}*{PATE} & 2.04 & 98.0\% & \multirow{2}*{99.20\%} & 1.21\% \\
            & {} & 8.03 & 98.1\% & {} & 1.11\% \\
            \cline{2-6}
            & \multirow{2}*{Proposed} & 1.93 & 98.52\% & \multirow{2}*{99.40\%} & 0.89\% \\
            & {} & 8.00 & 99.20\% & {} & 0.20\% \\
            \hline\hline
            \multirow{4}*{SVHN} & \multirow{2}*{PATE}  & 5.04 & 82.70\% & \multirow{2}*{92.80\%} & 10.88\% \\
            & {} & 8.19 & 90.70\% & {} & 2.26\% \\
            %& Improved PATE & 4.96 & 91.60\% & {} & 1.29\% \\ 
            \cline{2-6}
            & \multirow{2}*{Proposed} & 5.02 & 93.12\% & \multirow{2}*{95.30\%} & 2.29\% \\
            & {} & 8.18 & 93.71\% & {} & 1.68\% \\
            \hline\hline
             \multirow{4}*{CIFAR} & \multirow{2}*{RONA}  & 4.20 &  78.6 \% & \multirow{2}*{86.35\%} & 8.98\% \\
            & {} & 8.87 & 81.69\% & {} & 5.40\% \\
            \cline{2-6}
            & \multirow{2}*{Proposed}  & 4.21 &   81.76\% & \multirow{2}*{89.40\%} & 8.54\% \\
            & {} & 8.81 & 84.79\% & {} & 5.16\%  \\
			\hline
        \end{tabular}\vspace{-0.5cm}
    \end{center}
\end{table}

\section{CONCLUSION}
%To transfer knowledge from sensitive training data, t
This paper presented a three-player (teacher-student-discriminator) framework that transfers private knowledge and improves privacy and utility trade-off. The proposed framework combines KD from a teacher with GAN involving a student and a discriminator. A key insight in the proposed learning strategy is that the combination of KD and GAN provides additional quality supervision in addition to the distilled knowledge. Then, with a differential privacy mechanism, the proposed framework can establish a precise privacy guarantee of the training procedure while the training convergence can be quickly achieved even with limited training epochs and unlabelled training instances. Experimental results show that the proposed framework can act as the baseline framework for private knowledge transfer and achieve excellent utility and privacy trade-off on the datasets of MNIST, SVHN and CIFAR-10 for multi-label classification tasks. 

\section{ACKNOWLEDGEMENTS}
 This work was supported in part by the National Key R\&D Program of China (Grant No. 2018YFE0126300) and the National Science Foundation of China (Grant No. 61974133). 
 
%Since this transferring technique has exhibited a promising applicability on balancing utility and privacy. Future work may investigate the differential privacy mechanism for elaborately design the noise added and carefully track the privacy loss. 

%\ack We would like to thank the referees for their comments, which helped improve this paper considerably

%\bibliography{ecai}

\begin{thebibliography}{10}

\bibitem{abadi2016deep}
Martin Abadi, Andy Chu, Ian Goodfellow, H~Brendan McMahan, Ilya Mironov, Kunal
  Talwar, and Li~Zhang, `Deep learning with differential privacy', in {\em
  Proceedings of the 2016 ACM SIGSAC Conference on Computer and Communications
  Security}, pp. 308--318. ACM, (2016).

\bibitem{dwork2011differential}
Cynthia Dwork, `Differential privacy', {\em Encyclopedia of Cryptography and
  Security},  338--340, (2011).

\bibitem{dwork2006calibrating}
Cynthia Dwork, Frank McSherry, Kobbi Nissim, and Adam Smith, `Calibrating noise
  to sensitivity in private data analysis', in {\em Theory of cryptography
  conference}, pp. 265--284. Springer, (2006).

\bibitem{fredrikson2015model}
Matt Fredrikson, Somesh Jha, and Thomas Ristenpart, `Model inversion attacks
  that exploit confidence information and basic countermeasures', in {\em
  Proceedings of the 22nd ACM SIGSAC Conference on Computer and Communications
  Security}, pp. 1322--1333. ACM, (2015).

\bibitem{isola2017image}
Phillip Isola, Jun-Yan Zhu, Tinghui Zhou, and Alexei~A Efros, `Image-to-image
  translation with conditional adversarial networks', in {\em Proceedings of
  the IEEE conference on computer vision and pattern recognition}, pp.
  1125--1134, (2017).

\bibitem{krizhevsky2009learning}
Alex Krizhevsky, Geoffrey Hinton, et~al., `Learning multiple layers of features
  from tiny images', Technical report, Citeseer, (2009).

\bibitem{lecun1998gradient}
Yann LeCun, L{\'e}on Bottou, Yoshua Bengio, Patrick Haffner, et~al.,
  `Gradient-based learning applied to document recognition', {\em Proceedings
  of the IEEE}, {\bf 86}(11),  2278--2324, (1998).

\bibitem{lopez2015unifying}
David Lopez-Paz, L{\'e}on Bottou, Bernhard Sch{\"o}lkopf, and Vladimir Vapnik,
  `Unifying distillation and privileged information', {\em arXiv preprint
  arXiv:1511.03643}, (2015).

\bibitem{maddison2014sampling}
Chris~J Maddison, Daniel Tarlow, and Tom Minka, `A* sampling', in {\em Advances
  in Neural Information Processing Systems}, pp. 3086--3094, (2014).

\bibitem{mcmahan2018general}
H~Brendan McMahan and Galen Andrew, `A general approach to adding differential
  privacy to iterative training procedures', {\em arXiv preprint
  arXiv:1812.06210}, (2018).

\bibitem{mcmahan2017learning}
H~Brendan McMahan, Daniel Ramage, Kunal Talwar, and Li~Zhang, `Learning
  differentially private recurrent language models', {\em arXiv preprint
  arXiv:1710.06963}, (2018).

\bibitem{mironov2017renyi}
Ilya Mironov, `R{\'e}nyi differential privacy', in {\em 2017 IEEE 30th Computer
  Security Foundations Symposium (CSF)}, pp. 263--275. IEEE, (2017).

\bibitem{mironov2019r}
Ilya Mironov, Kunal Talwar, and Li~Zhang, `R$\backslash$'enyi differential
  privacy of the sampled gaussian mechanism', {\em arXiv preprint
  arXiv:1908.10530}, (2019).

\bibitem{netzer2011reading}
Yuval Netzer, Tao Wang, Adam Coates, Alessandro Bissacco, Bo~Wu, and Andrew~Y
  Ng, `Reading digits in natural images with unsupervised feature learning',
  (2011).

\bibitem{papernot2016semi}
Nicolas Papernot, Mart{\'\i}n Abadi, Ulfar Erlingsson, Ian Goodfellow, and
  Kunal Talwar, `Semi-supervised knowledge transfer for deep learning from
  private training data', {\em arXiv preprint arXiv:1610.05755}, (2016).

\bibitem{papernot2018scalable}
Nicolas Papernot, Shuang Song, Ilya Mironov, Ananth Raghunathan, Kunal Talwar,
  and {\'U}lfar Erlingsson, `Scalable private learning with pate', {\em arXiv
  preprint arXiv:1802.08908}, (2018).

\bibitem{romero2014fitnets}
Adriana Romero, Nicolas Ballas, Samira~Ebrahimi Kahou, Antoine Chassang, Carlo
  Gatta, and Yoshua Bengio, `Fitnets: Hints for thin deep nets', {\em arXiv
  preprint arXiv:1412.6550}, (2014).

\bibitem{shokri2015privacy}
Reza Shokri and Vitaly Shmatikov, `Privacy-preserving deep learning', in {\em
  Proceedings of the 22nd ACM SIGSAC conference on computer and communications
  security}, pp. 1310--1321. ACM, (2015).

\bibitem{shokri2017membership}
Reza Shokri, Marco Stronati, Congzheng Song, and Vitaly Shmatikov, `Membership
  inference attacks against machine learning models', in {\em 2017 IEEE
  Symposium on Security and Privacy (SP)}, pp. 3--18. IEEE, (2017).

\bibitem{song2017machine}
Congzheng Song, Thomas Ristenpart, and Vitaly Shmatikov, `Machine learning
  models that remember too much', in {\em Proceedings of the 2017 ACM SIGSAC
  Conference on Computer and Communications Security}, pp. 587--601. ACM,
  (2017).

\bibitem{tramer2016stealing}
Florian Tram{\`e}r, Fan Zhang, Ari Juels, Michael~K Reiter, and Thomas
  Ristenpart, `Stealing machine learning models via prediction apis', in {\em
  25th $\{$USENIX$\}$ Security Symposium ($\{$USENIX$\}$ Security 16)}, pp.
  601--618, (2016).

\bibitem{wang2019private}
Ji~Wang, Weidong Bao, Lichao Sun, Xiaomin Zhu, Bokai Cao, and S~Yu Philip,
  `Private model compression via knowledge distillation', in {\em Proceedings
  of the AAAI Conference on Artificial Intelligence}, volume~33, pp.
  1190--1197, (2019).

\bibitem{wang2018kdgan}
Xiaojie Wang, Rui Zhang, Yu~Sun, and Jianzhong Qi, `Kdgan: knowledge
  distillation with generative adversarial networks', in {\em Advances in
  Neural Information Processing Systems}, pp. 775--786, (2018).

\bibitem{xu2017training}
Zheng Xu, Yen-Chang Hsu, and Jiawei Huang, `Training shallow and thin networks
  for acceleration via knowledge distillation with conditional adversarial
  networks', {\em arXiv preprint arXiv:1709.00513}, (2017).

\bibitem{yu2017seqgan}
Lantao Yu, Weinan Zhang, Jun Wang, and Yong Yu, `Seqgan: Sequence generative
  adversarial nets with policy gradient', in {\em Thirty-First AAAI Conference
  on Artificial Intelligence}, (2017).

\bibitem{yu2019differentially}
Lei Yu, Ling Liu, Calton Pu, Mehmet~Emre Gursoy, and Stacey Truex,
  `Differentially private model publishing for deep learning', {\em arXiv
  preprint arXiv:1904.02200}, (2019).

\bibitem{zhang2017adversarial}
Yizhe Zhang, Zhe Gan, Kai Fan, Zhi Chen, Ricardo Henao, Dinghan Shen, and
  Lawrence Carin, `Adversarial feature matching for text generation', in {\em
  Proceedings of the 34th International Conference on Machine Learning-Volume
  70}, pp. 4006--4015. JMLR. org, (2017).

\end{thebibliography}

\end{document}